\documentclass[twocolumn,showpacs,preprintnumbers]{revtex4}
\usepackage{graphicx}
\usepackage{dcolumn}
\usepackage{bm}

\draft

\begin{document}

\title{Pressure-induced lattice instabilities and superconductivity in YBCO}

\author{M. Calamiotou$^1$, A. Gantis$^1$, E. Siranidi$^2$, D. Lampakis$^2$, J. Karpinski$^3$, and E. Liarokapis$^2$}

\address{$^1$Solid State Physics Department, School of Physics, University of Athens, GR-15784 Athens, Greece}
\address{$^2$Department of Physics, National Technical University of Athens, 157 80 Athens, Greece}
\address{$^3$Laboratory for Solid State Physics, ETH, CH-8093 Z$\ddot{u}$rich, Switzerland}

\begin{abstract}
Combined synchrotron angle-dispersive powder diffraction and
micro-Raman spectroscopy are used to investigate pressure-induced
lattice instabilities that are accompanied by superconducting
T$_{\rm c}$ anomalies in YBa$_{\rm 2}$Cu$_{\rm 4}$O$_{\rm 8}$ and
optimally doped YBa$_{\rm 2}$Cu$_{\rm 3}$O$_{\rm 7-\delta}$, in
comparison with the non-superconducting PrBa$_{\rm 2}$Cu$_{\rm
3}$O$_{\rm 6.92}$. In the first two superconducting systems there
is a clear anomaly and hysteresis in the evolution of the lattice
parameters and increasing lattice disorder with pressure, which
starts at $\approx3.7$ GPa. On the contrary, in the Pr-compound
the lattice parameters follow very well the expected equation of
state (EOS) up to 7 GPa. The micro-Raman data of the
superconducting compounds show that the energy and width of the
A$_{\rm g}$ phonons exhibit anomalies over the same pressure range
where the lattice parameters deviate from the EOS and the average
Cu2-O$_{pl}$ bond length exhibits a strong contraction that
correlates with the non-linear pressure dependence of T$_{\rm c}$.
The anomalous Raman behavior is not observed  for the non
superconducting Pr compound, clearly indicating a connection with
the charge carriers. It appears that the cuprates close to optimal
doping are at the edge of lattice instability.
\end{abstract}


\maketitle

\subsection*{Introduction}

It is well accepted that structural and electronic inhomogeneities
constitute intrinsic properties of cuprate superconductors \cite
{Science1,Science2,Lampakis06}. To this context the study of any
lattice distortions induced by application of either internal
chemical or external hydrostatic pressure \cite
{Gantis,Calamiotou1,Calamiotou} that modify the transition
temperature (T$_{\rm c}$) is important for understanding the role
of lattice effects in the high T$_{\rm c}$ superconductivity.
Lattice instabilities in hydrostically compressed
YBa$_{2}$Cu$_{3}$O$_{\rm y}$ (Y123) and YBa$_{2}$Cu$_{4}$O$_{8}$
(Y124) cuprates where T$_{\rm c}$ dependence on pressure shows
saturation or non-linear behavior \cite {Koch,Scholtz}, manifest
themselves in the Raman phonon frequencies as a deviation from an
expected linear behavior at a critical pressure range 2.5-6 GPa
\cite {Liarokapis,Lampakis1}. Recent structural investigations,
using synchrotron angle-dispersive powder diffraction and dense
sampling on optimally doped Y123 superconductor, have revealed in
the pressure range 3.7 GPa$<$p$<$10 GPa a clear deviation of the
lattice constants from the expected EOS, strong hysteresis
effects, and the formation of an additional new textured phase
\cite {Calamiotou}. Interatomic distances in the unit cell of Y123
such as the Ba distance from the basal plane, the Cu2-O$_{\rm pl}$
bond length, and the Cu2-Cu1 distance along the c-axis have been
found to exhibit also a non linear evolution with applied pressure
that correlates with modifications of the Raman spectra of the
in-phase O$_{\rm pl}$ and the O$_{\rm ap}$ phonon modes \cite
{Calamiotou}. The correlation of the structural characteristics
with the Raman frequency modifications \cite {Calamiotou} and
corresponding changes of T$_{\rm c}$ \cite {Koch} imply that the
trigger of the lattice instabilities lies among the CuO$_{\rm 2}$
and BaO planes. Charge redistribution among the planes, that has
been found to occur in the case of chemical pressure \cite
{Karpinski}, also affects the T$_{\rm c}$, and must certainly play
a role in the observed effects.

The main questions that arise is whether similar effects are
present in other cuprates and how they relate to the amount of
charge carriers and superconductivity. The isostructural to Y123
compound PrBa$_{\rm 2}$Cu$_{\rm 3}$O$_{\rm 7}$ (Pr123) is a good
example to study, having apparently no free carriers
\cite{Calamiotou1}. Structural and Raman studies under hydrostatic
pressure in this compound are absent. The Y124 superconductor is
another good case study, since it can be considered as a model
compound for the underdoped region of the YBCO system. Previous
structural studies under hydrostatic pressure up to 5 GPa (using a
laboratory source) gave only some marginal evidence of a
non-linear pressure dependence above 4 GPa \cite {NELMES}. We
present here high quality synchrotron angle-dispersive powder
diffraction data and micro-Raman spectra under hydrostatic
pressure for the Pr123 (T$_{\rm c}$=0K) and Y124 (T$_{\rm c}$=80K)
compounds in comparison with the corresponding ones for Y123
(T$_{\rm c}$=92K) \cite{Calamiotou}. The pressure region up to 13
GPa has been investigated with a pressure step of $\approx0.5$
GPa. Our combined structural and spectroscopic results provide
strong evidence that the presence of charge carriers and the
pressure-induced lattice instabilities in the superconducting
cuprates are strongly related, thus affecting the T$_{\rm c}$
dependence on pressure.

\subsection*{Experimental details}

We have studied a powder PrBa$_{\rm 2}$Cu$_{\rm3}$O$_{\rm 6.92}$
sample with an oxygen content comparable to the previously
investigated optimally doped Y123 \cite {Calamiotou} and the
YBa$_{\rm 2}$Cu$_{\rm4}$O$_{\rm 8}$ compounds. Synchrotron
angle-dispersive powder diffraction experiments at high pressures
have been carried out at the Swiss-Norvegian beamline BM01A of
ESRF (Grenoble), using a diamond anvil cell (DAC) and a 4:1
methanol-ethanol mixture as pressure medium, which is known to
remain hydrostatic and do not cause peak broadening in powder
samples at least up to $\sim$10 GPa \cite {Kenichi}. Ruby crystals
were used for measuring the pressure and monitoring any
inhomogeneity in the pressure distribution. Diffraction patterns
have been collected with a wavelength $\lambda$=0.7117$\AA$ and a
MAR345 image plate detector. The pressure cell was left to relax
for approx. 20-30 min at each pressure. The 2D raw images have
been converted to 2$\theta$ patterns, after correcting for
distortions and refining the detector-sample distance using a
LaB$_{6}$ standard by the program FIT-2D \cite {Hammersley}.

The Raman spectra were obtained at room temperature and high
hydrostatic pressures with a T64000 Jobin–Yvon triple spectrometer
equipped with a liquid nitrogen cooled charge coupled device (CCD)
and a microscope lens of magnification $\times$40. A Merrill
Bassett type diamond anvil cell (DAC) was used for the
high-pressure measurements (up to 6.8 GPa), which allowed the
Raman studies to be carried out in a back scattering geometry. The
pressure-transmitting medium was a mixture of methanol-ethanol,
with analogy (4:1). Silicon single crystals distributed around the
sample were used for calibration of the pressure and for
monitoring the hydrostatic conditions. The 514.5 nm line from an
Ar$^{+}$ laser was used for excitation, which produced the lowest
luminescence from the diamonds. The laser beam was focused on the
sample at a spot of diameter 4 $\mu$m while the power level was
kept below 0.20 mW. Typical accumulation times were up to 3 hours
depending on the scattering polarization.

\subsection*{Results}

\subsection{\emph{Average structure}}

Structural data have been obtained by analyzing the
intensity-vs-2$\theta$ diffraction patterns with the Rietveld
method using Fullprof \cite {Rodriguez-Carvajal} and a
pseudo-Voigt line profile function. The diffraction patterns of
Pr123 have been refined with the orthorhombic Pmmm space group in
the whole pressure region. Minor impurity lines belonging to
BaCuO$_{2}$ \cite {Calamiotou1} have been excluded. Lattice
constants for Y124 have been refined with the Ammm space group
using the LeBail method, implemented in Fullprof \cite
{Rodriguez-Carvajal}. Weak peaks appearing for $\emph{p}\geq
3.7GPa$ have been excluded as discussed below. Fig.1a shows a
characteristic synchrotron diffraction pattern of Pr123, at
$\emph{p}=4.7$ GPa, together with the results of the Rietveld
refinement and Fig.1b the corresponding pattern of Y124 with the
results of LeBail refinement. The pressure dependence of the a-,
b-axis is shown in Fig.2a for Pr123, Fig.2b for Y123 (from
ref.\cite {Calamiotou}), and Fig.2c for Y124. The evolution of the
c-axis with the applied hydrostatic pressure for Pr123 is shown in
Fig.3a in comparison with the corresponding one for the optimally
doped Y123 (Fig.3b) from ref.\cite {Calamiotou}, whereas the
results for Y124 are presented in Fig.3c. Full symbols represent
the values for increasing the pressure and open symbols those for
releasing the pressure. In the case of Y124, we have also carried
out measurements while increasing again the pressure after
pressure release (full triangles in Fig.2c and Figs.3c, 3f). The
lattice constants have been fitted to the empirical Murnaghan
equation of state $q=q_o [ p(\kappa_q /\kappa_q ^\prime
)+1]^{-\kappa_q^\prime}$, where $\emph{p}$ is the applied
pressure, $\emph{q}$ is the lattice constant under compression,
$\kappa_{q}$ its compressibility and $\kappa_{q} ^\prime$ a
parameter that expresses the pressure dependence of the
compressibility. For Pr123 the solid lines are fits to the data
for increasing pressure up to 7 GPa. For Y123 and Y124 the solid
lines are fits to the data for increasing pressure where the
pressure region 3.7$<\emph{p}<$8 GPa has been excluded as it will
be discussed below. The values of $\kappa_{a}=2.6\times 10^{-3}$
GPa$^{-1}$, $\kappa_{b}=2.1\times 10^{-3}$ GPa$^{-1}$ and
$\kappa_{c}=4.1\times 10^{-3}$ GPa$^{-1}$ obtained for Pr123 are
slightly smaller but comparable to those of Y123
\cite{Calamiotou}. The Y124 compound exhibits strong anisotropy in
all three lattice directions with corresponding compressibilities
$\kappa_{a}=3.7\times 10^{-3}$ GPa$^{-1}$, $\kappa_{b}=1.9\times
10^{-3}$ GPa$^{-1}$ and $\kappa_{c}=4.5\times 10^{-3}$ GPa$^{-1}$.
Upon decompression the c axis evolution of the Y124 compound can
be best described by the dashed line (Fig.3c), which is a
Murnaghan equation with $\kappa_{c}=4.2\times 10^{-3}$ GPa$^{-1}$
and a much bigger $\kappa_{c} ^\prime$ value compared to the
corresponding one for pressure increase.

For Pr123 the evolution of all lattice constants with increasing
pressure follows very well the Murnaghan equation of state up to
$\sim$7 GPa. Upon pressure release the effect of pressure on the
unit cell is reversible, i.e. the values of lattice constants
coincide with those at increasing pressure. This is in contrast
with the evolution of the c-axis of the Y123 compound. The c-axis
exhibits a strong deviation from the expected equation of state
that starts around 3.7 GPa and extends to 8-10 GPa, and there is a
clear irreversibility of the effect of pressure, which results in
a strong hysteresis \cite{Calamiotou}. The c-axis of Y124 shows
also a deviation from EOS in the same pressure region, however the
effect is less intense than in Y123.

The a- and b-axis of Y123 deviate also from the EOS
\cite{Calamiotou}, but to a smaller degree than the c-axis
(Fig.2b). For Pr123 no deviation is detected (Fig.2a), while for
Y124 the deviation in the b- and a-axis is more pronounced than in
Y123, with the b-axis showing a larger deviation than the a-axis
(Fig.2c). It is worth to mention that when pressure was increased
again in Y124 after pressure release, the values of the c-axis are
placed between the data of pressure increase and pressure release,
but closer to the second data (Fig.3c). In the case of the a- and
b-axis, the increase of pressure for a second time brings the data
closer to those of pressure release for the a-axis and of the
original pressure increase for the b-axis. However, the observed
modifications in those axes are much smaller than for c-axis and
the effect might be related to the anisotropicity of the compound.
For Pr123 and Y123 we have no data with increasing pressure for a
second time to compare, but as for Pr123 there is no deviation
from the EOS and the data of pressure increase or release
coincide, we could not observe any changes. Furthermore, in Y123
the deviations from the EOS for the a- and b-axis are smaller than
in Y124 and any changes, even if present, might not be able to be
detected.

\subsection{\emph{Microstructure}}

With increasing pressure the width of the diffraction peaks is
increasing. In the case of Y123, the Williamson-Hall plots have
revealed that the application of pressure induces a
microstrain-type diffraction peak broadening while size-type
broadening is negligibly small \cite {Calamiotou}. Although it is
very difficult to assess the origin of microstrains,
pressure-induced disorder is most likely associated with the
observed line broadening in the pressure region up to 10 GPa. In
order to examine any correlation of pressure-induced disorder to
the observed deviations from the expected EOS, we have estimated
the upper limit of apparent microstains-disorder in all compounds.
Since we are interested to investigate their evolution with
pressure rather than absolute values, we have used the width of
the single, non overlapping 113 Bragg peak. The width of the 113
reflection is not affected by intrinsic defects, such as
dislocations and stacking faults, as we have found in the case of
Pr123 \cite{Calamiotou1}. The 113 peaks have been fitted with a
pseudo-Voigt function and the apparent microstrain, $\varepsilon$,
has been calculated as $\varepsilon=(W-W_{i})/tan\theta$, where
$W$ and $W_{i}$ are the full widths at half maximum of the sample
and the standard LaB$_{6}$ respectively, that has been used to
determine the instrumental resolution.

The evolution of microstrains-disorder with pressure for Pr123,
Y123 and Y124 is shown in Fig.3d, 3e and 3f respectively.  Full
symbols correspond to the values for increasing the pressure and
open symbols to those for releasing the pressure. Dotted lines are
guide to the eye to emphasize the regions with different increase
of disorder upon pressure. Fig.3d shows that microstrains in Pr123
remain almost constant up to a pressure of 7 GPa, starting to
increase exactly at the same pressure where the c-axis deviates
from the expected equation of state. \emph{No irreversibility and
hysteresis effects are observed}. On the contrary, the
microstrains of Y123 start to increase at $\emph{p}\sim$ 3.7 GPa
where the c-axis deviates from the equation of state, for
$\emph{p}>$7 GPa they continue to increase with a bigger slope and
for $\emph{p}>$10 GPa they increase further. In addition
\emph{pronounced hysteresis effects are present} (Fig.3e).
Analogous results have been obtained for Y124 (Fig.3f). It is
clear  from Fig.3 that the increase of microstrains occurs in all
cases exactly at the same pressure where there is an anomaly in
the expected lattice constants behavior on pressure. With the
reduction of pressure to ambient conditions, the Pr123 compound
practically returns to the original state of microstrains
indicating that they are not permanent lattice distortions. On the
contrary, the Y123 and Y124 compounds show hysteresis in the whole
pressure range. It should be noted that, the 2D patterns obtained
with the $\emph{cake}$ function of the FIT2D \cite{Hammersley}
software do not bear any sign of a non-hydrostatic pressure on the
diffraction lines of Pr123,Y123 (Fig.4 in ref.\cite {Calamiotou})
and  Y124. This is in agreement with all measurements performed up
to now that do not provide any evidence for a non hydrostatic
environment in the methanol-ethanol mixture for pressures up to 10
GPa.

\subsection{\emph{Raman spectra}}

Typical Raman spectra of the hydrostatically compressed Pr123
compound are shown in Fig.4. All five A$_{\rm g}$ symmetry phonons
with eigenvectors along the c-axis were studied in the two
scattering polarizations, xx (Fig.4a) and zz (Fig.4b)
corresponding to incident and scattered light polarization
parallel or perpendicular to the CuO$_2$ planes. Those are at
$\sim$128 cm$^{-1}$ (Ba-atom), $\sim$150 cm$^{-1}$ (Cu$_{\rm
pl}$-atom), $\sim$300 cm$^{-1}$ (out of phase (B$_{\rm 1g}$-like)
vibrations of the O$_{\rm pl}$ atoms), $\sim$435 cm$^{-1}$ (in
phase vibrations of the O$_{\rm pl}$ atoms) and $\sim$523
cm$^{-1}$ (vibrations of the O$_{\rm ap}$ atoms). Fig.5 shows that
the energy of the  four A$_{\rm g}$ symmetry phonons in Pr123
increases almost linearly with pressure. For comparison, the
pressure dependence of the ortho-I and ortho-II phases of Y123,
corresponding to oxygen content $\approx$7 (Y123) and $\approx$6.5
(Y1236.5), are also shown, where it is clear that in the region
2.5-4 GPa there is no increase in the phonon energy despite the
increase in pressure. \emph{This softening of the four A$_{\rm g}$
phonons is in contrast to their almost linear behavior for Pr123.}
For the B$_{\rm 1g}$-like mode the phonon energy has been found to
increase almost linearly with increasing pressure, as expected,
without showing any anomaly, for all compounds.

Figure 6 presents the variation of the corresponding phonon width
of the four A$_{\rm g}$ symmetry phonons for the sets of
compounds, which again indicates that both Y123 and Y1236.5 have
an unusual behavior with pressure compared with Pr123. Especially
the apical phonon mode shows a considerable increase in width for
p$\approx$2 GPa and then a decrease to almost the original value
for p$\approx$3.7 GPa followed by an increase. This unconventional
variation of the width with pressure is not a result of error in
the fitting procedure, as it is quite clear from the spectra
(Fig.1 of ref. \cite {Liarokapis}). The initial increase in width
and frequency of the mode up to $\approx$2 GPa is a typical
increase of the modes with pressure. Afterwards, both the width
and the frequency show abnormal behavior, which indicates that
another mechanism is active. This could be a pressure induced
lattice instability that results in phase separation, mode
frequency softening, and decrease in the width, which finally
relaxes the internal strains.

\subsection*{Discussion}

The evolution of microstrains-disorder in Pr123 indicates two
regions of almost linear pressure dependence for the compound;
namely for pressures up to 7 GPa and above that (Fig.3d). The
increase of disorder for 7$<\emph{p}\leq$10 GPa which correlates
with the deviation of the c-axis values from the expected EOS
(Fig.3a) would imply that the system undergoes a phase transition
likely related to a distorted orthorhombic phase. In the case of
Y123 and Y124 additional features are evident in the
microstrains-disorder evolution on pressure. There is an
intermediate pressure region (3.7-7 GPa), where microstrains
exhibit again an almost linear increase on pressure but with a
different slope (Fig.3e and 3f). Interestingly, in the two
pressure regions, i.e., $\emph{p}<3.7GPa$ and 7$<\emph{p}\leq$10
GPa microstrains increase with the same pressure slopes in the
Pr123, Y123 and Y124 compounds. Since Pr123 and Y123 have the same
structure and comparable compressibilities, the increased
microstrains-disorder in the intermediate (3.7-7 GPa) pressure
region (observed under the same experimental conditions only in
Y123 and Y124) implies that pressure-induced disorder in that
region is an intrinsic effect of Y123 and Y124. It should be
mentioned that at the pressure of $\approx$3.7 GPa four new weak
Bragg peaks appear in the diffraction patterns of both Y123 (Fig.3
in ref.\cite {Calamiotou}) and Y124 (Fig.1b), which are not
observed in Pr123 (Fig.1a). The new weak lines could not be due to
the formation of a superstructure, as the systematic investigation
of all observed lines has proved. Moreover the 2D diffraction
images revealed that these new lines exhibit strong texture and
strain effects unlikely to those of the main phase (Fig.4 in
ref.\cite {Calamiotou}). We therefore assume that they are due to
the formation of an additional textured and distorted phase at
this critical pressure in both Y123 and Y124 compounds. The new
weak lines disappear completely upon pressure release below
roughly the same pressure ($\approx$3 GPa)(Fig.3 in
ref.\cite{Calamiotou}). When pressure is increased again in Y124
after pressure release they appear again however their intensity
is strongly reduced implying a correlation of the presence of the
new phase with the corresponding c-axis anomaly (Fig.3c).

The increased disorder in both Y124 and Y123 compounds over the
intermediate (3.7-7 GPa) pressure region appears together with the
hysteresis, the anomaly in the lattice constants evolution (Fig.3)
and the new lines reminiscent of the development of a new phase.
In the diffraction pattern of Y124, we have also observed at 10.3
GPa the appearance of two additional lines at 2$\theta$ $\approx$
15.5 and 28 degrees, with texture effects, which disappear again
for $\emph{p}<9 GPa$ on decompression. At the same pressure the
b-axis deviates from the expected equation of state. The formation
of the new phases happens at the pressure where there is a
deviation from the EOS, and whenever this does not occur, as in
Pr123, the new lines do not appear. Apparently, they are related
with a phase separation, which is triggered by the hydrostatic
pressure and causes deviations from the EOS. To this context the
different microstrain regions marked by dotted straight lines with
a different pressure slope in Figures 3(e) and 3(f) may indicate
the formation of different distorted orthorhombic micro-phases in
Y123 and Y124. All these effects are absent in Pr123 up to a
pressure of 7 GPa.

For the in-phase mode the high-pressure Raman results for the Y123
compound have indicated modifications, which appear as a double
peak at about 2 GPa \cite {Lampakis}, i.e. at the pressure where
the bond distance Cu2-O$_{\rm pl}$ and the position of the Ba atom
show a non-linear dependence on pressure \cite {Calamiotou}.
Similar modifications from linearity with increasing pressure were
also detected for the apical oxygen frequency \cite {Lampakis},
which correlates (as expected) with the pressure dependence of the
Cu2-Cu1 bond lengths \cite {Calamiotou}. The same Raman results
have also been obtained for the Y124 compound \cite {Lampakis,
Lampakis1}. However, the non superconducting Pr123 compound does
not show such anomalies. In complete agreement with the
corresponding Raman results, the average Cu2-O$_{\rm pl}$ bond
length obtained from the Rietveld refinement for Pr123 (shown in
Fig.7a) exhibits only a very small, if any, modification at the
characteristic pressure of $\approx$ 3.7 GPa contrary to the
results of the Y123 compound (Fig.7b). The abrupt contraction of
the Cu2-O$_{\rm pl}$ bond in Y123 at $\approx$ 3.7 GPa may be
related \emph{to a pressure-induced charge redistribution}.

The combined XRD and Raman results show for Y123 and Y124 in the
pressure region 3.7$<$p$<$8 GPa; a clear deviation from the
expected EOS of the c-axis (Figs.3b, 3c),  modifications of
disorder (Figs.3e, 3f), which correlate with a softening of the
four A$_{\rm g}$ phonons (Fig.5) and  anomaly in their width
(Fig.6). Furthermore, there is also a strong irreversibility and
hysteresis together with the appearance of a new phase. In the
corresponding data of the isostructural Pr123 no similar phenomena
are observed. One could assume that for 3.7$<\emph{p}<8$ GPa a new
phase is created with pressure in these cuprates, which is
accompanied by modifications in the lattice along the Cu2-O$_{\rm
pl}$ and Cu2-O$_{\rm ap}$ bonds, directly affecting the electronic
states, the carrier distribution, and the transition temperature
in Y123 and Y124, while having no appreciable difference in Pr123,
which apparently does not have carriers \cite{Calamiotou1}. The
main effects are observed for the A$_{\rm g}$ symmetry phonons
with eigenvectors along the c-axis while the B$_{\rm 1g}$ phonons
are not affected. The pressure data in the cuprates studied point
to a lattice instability close to optimal doping, which once
exceeded by pressure (present work) or doping \cite{Kaldis},
induces local lattice distortions mostly in the CuO$_{\rm 2}$
planes that modify the transition temperature.

Previous studies in both Y123 and Y124 compounds have revealed
that at those critical pressures (3.7 GPa and 10.3 GPa) there are
modifications of the critical temperature T$_{\rm c}$ dependence
on pressure \cite {Koch,Scholtz}. Moreover irreversibility and
hysteresis were  observed in the pressure dependence of T$_{\rm
c}$ for the Y124 compound by Scholtz et al \cite{Scholtz}. It was
suggested that a possible transformation of the sample for
pressures above 20 GPa was the reason for this unexpected
hysteresis. However, our data and more precisely the modification
of the c-axis compressibility of Y123 and Y124 in the intermediate
(3.7-7 GPa) pressure region, which correlates with the increase of
microstrains, the appearance of extra Bragg peaks, and the
modifications in T$_{\rm c}$ implies that the origin of these
phenomena must be an intrinsic instability of the superconducting
cuprates, which appears at much lower hydrostatic pressures
($\sim$3.7 GPa), affects T$_{\rm c}$ and results in the
development of another phase that appears together with the
deviation from the normal EOS. Such effects and a new phase are
absent in the non-superconducting Pr123 at least up to 7 GPa.

\subsection*{Conclusions}

Combined synchrotron xrd and Raman measurements from
superconducting compounds Y123 and Y124 show pressure-induced
lattice anomalies (deviation from the normal EOS of the lattice
constants), irreversibility and hysteresis, disorder (estimated by
the line broadening), the appearance of additional diffraction
peaks apparently related with a new phase, softening and abnormal
pressure-dependence of the width for all A$_{\rm g}$-symmetry
phonon modes. These modifications fully correlate with anomalies
in the T$_{\rm c}$-dependence on pressure. The absence of similar
effects in the non-superconducting isostructural Pr123 is a strong
indication that the observed effects are related also with the
presence of free carriers. The anomalies in the Raman-active
phonon modes at the same pressures with the bond length changes
calculated from the synchrotron diffraction data point to a charge
redistribution within and around the CuO$_{\rm 2}$ planes. This
justifies the full correlation with the pressure modifications in
the transition temperature and the absence of similar effects in
Pr123. Furthermore, it provides a clue that close to optimal
doping the cuprates are at a lattice instability, which seems to
be an intrinsic property in the compounds studied. Although we
cannot anticipate the behavior of the rest of the cuprates, our
up-to-now data provide a hint about the role of the lattice in the
high T$_{\rm c}$ superconductivity of the cuprates.

\subsection*{Acknowledgements}
We thank the Swiss-Norwegian committee for beam time allocation at
the SNBL BM01A, at ESRF, and Prof. V. Dmitriev for discussion and
help during the experiments at BM01A. The team from NTUA
acknowledges support from CoMePhS (STRP project of E.U.). Partial
support from the Research Account of the University of Athens
(Kapodistrias) is acknowledged.

\bigskip

\subsection*{FIGURE CAPTIONS}

Figure 1. (Color online) Rietveld refinement of synchrotron powder
diffraction patterns for Pr123 at $\emph{p}$=4.7 GPa (a) and
LeBail refinement for Y124 at $\emph{p}$=4.7 GPa (b). Experimental
(circles), calculated (continuous line) intensities, their
difference (bottom line) and corresponding positions of Bragg
peaks (bars). The part of the pattern near 2$\theta$=20$^{o}$,
where the lines from the gasket appear, has been excluded.($*$)
shows the peaks of impurity BaCuO$_{2}$ in Pr123 and
($\diamondsuit$) the weak peaks appearing for $\emph{p}\geq
3.7GPa$ in Y124.

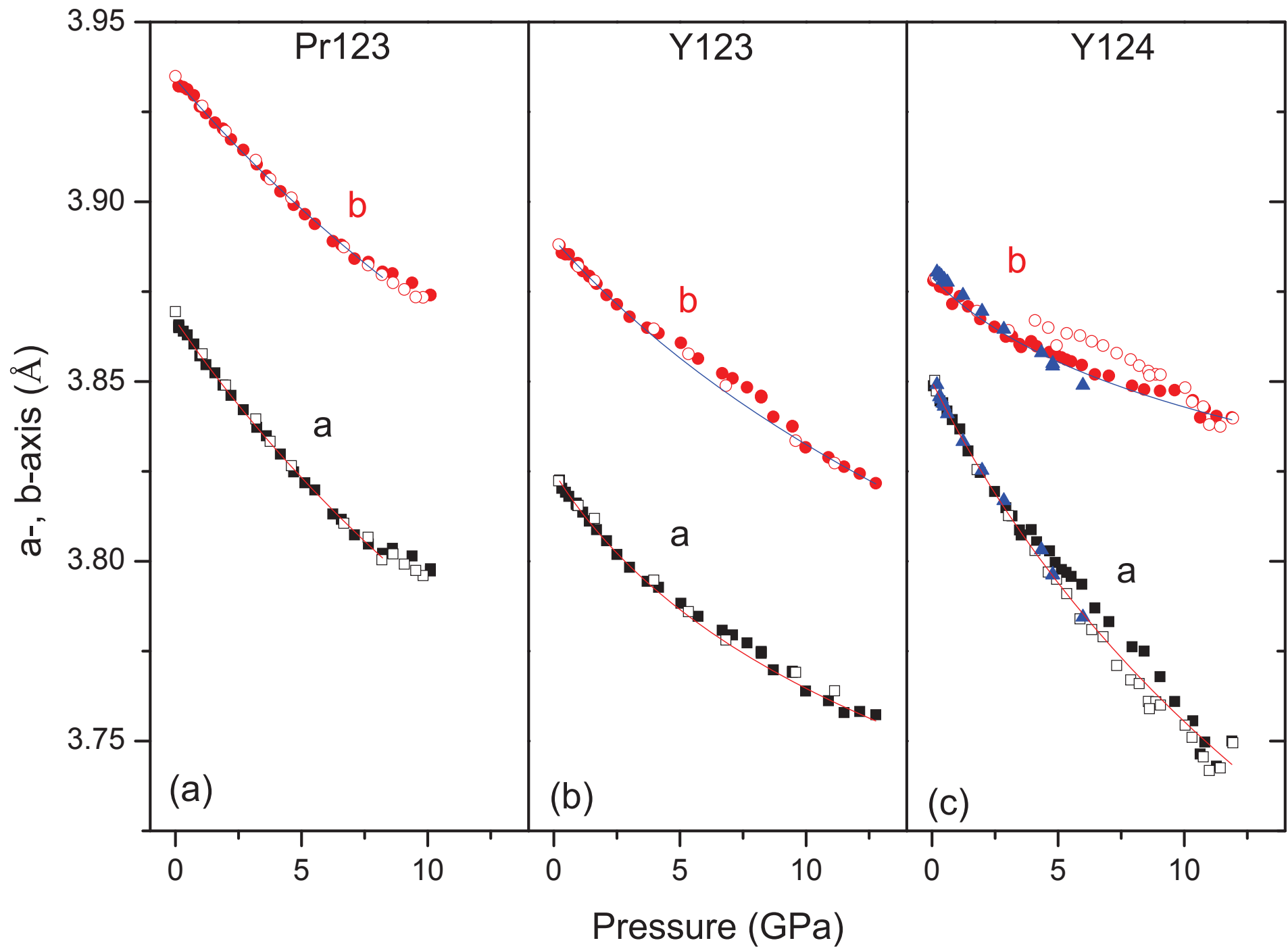
Figure 2. (Color online) Pressure dependence of the a-, b-axis for
the Pr123 (a), Y123 (ref. \cite{Calamiotou}) (b) and Y124 (c)
compounds. Full symbols correspond to increasing pressure and open
symbols to pressure release. Full triangles correspond to the
increase of pressure for second time. Solid lines are fits to the
Murnaghan EOS.
\bigskip

Figure 3. (Color online) Pressure dependence of the c-axis for the
Pr123 (a), Y123 ( from ref. \cite{Calamiotou}) (b) and Y124 (c)
compounds. Evolution of apparent disorder $\varepsilon$ with
applied pressure for the Pr123 (d), Y123 (e) and Y124 (f)
compounds. Full symbols correspond to increasing pressure and open
symbols to pressure release. Full triangles correspond to the
increase of pressure for second time. Solid lines (and dashed line
in 2c) are fits to the Murnaghan EOS as described in the text.
Dotted lines in 2d-f are guide to the eye to emphasize the regions
with different increase of disorder upon pressure.
\bigskip

Figure 4. Typical Raman spectra of the Pr123 compound at selected
pressures in the xx (a) and zz (b) scattering configuration.
\bigskip

Figure 5. (Color online) The variation of the energy of the
Cu$_{\rm pl}$ (a), O$_{\rm ap}$ (b), Ba (c) and O$_{\rm pl}$ (d)
phonon upon pressure for different cuprates; YBa$_{\rm 2}$Cu$_{\rm
3}$O$_{\rm 6.5}$ (Y1236.5), YBa$_{\rm 2}$Cu$_{\rm 3}$O$_{\rm 7}$
(Y123) and Pr123.
\bigskip

Figure 6. (Color online) The variation of the width of the
(Cu$_{\rm pl}$ (a), O$_{\rm ap}$ (b), Ba (c) and O$_{\rm pl}$ (d)
phonon upon pressure for the cuprates of Fig.5.
\bigskip

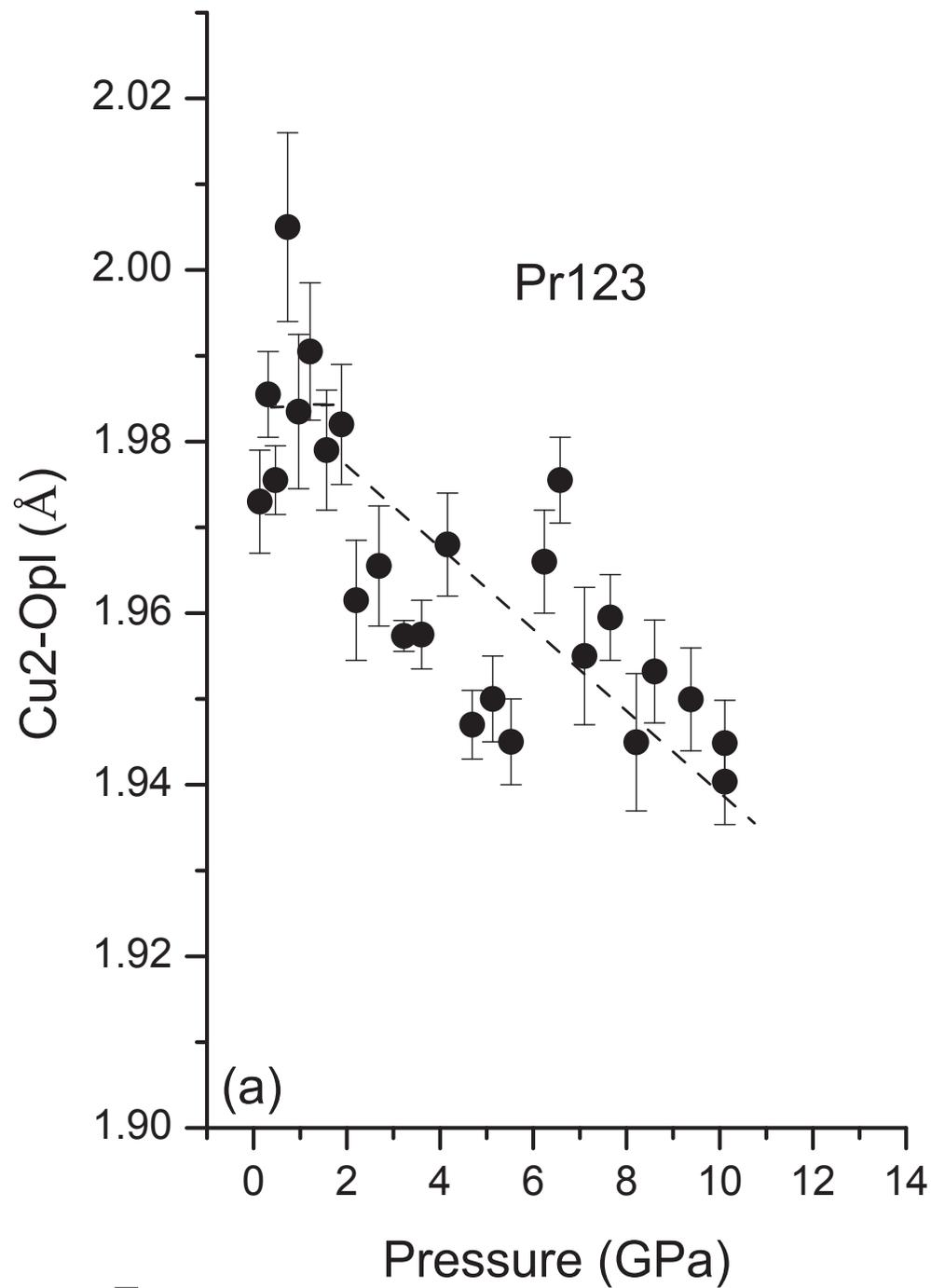
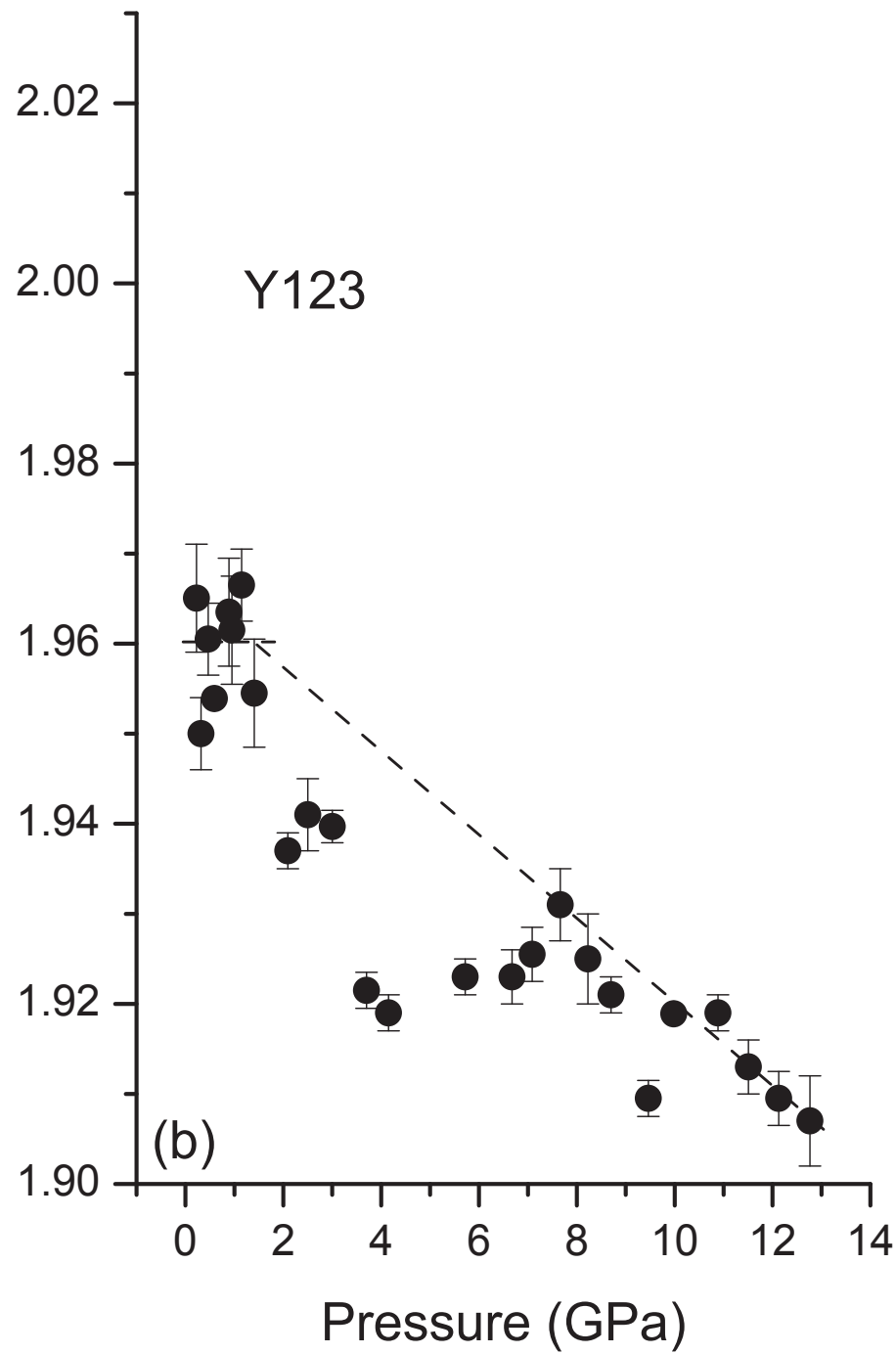
Figure 7.  The pressure dependence of the Cu2-O$_{\rm pl}$ bond
length for the  Pr123 (a) in comparison to that of Y123 (ref.
\cite{Calamiotou})(b). Dashed lines are guide to the eye.

\bigskip

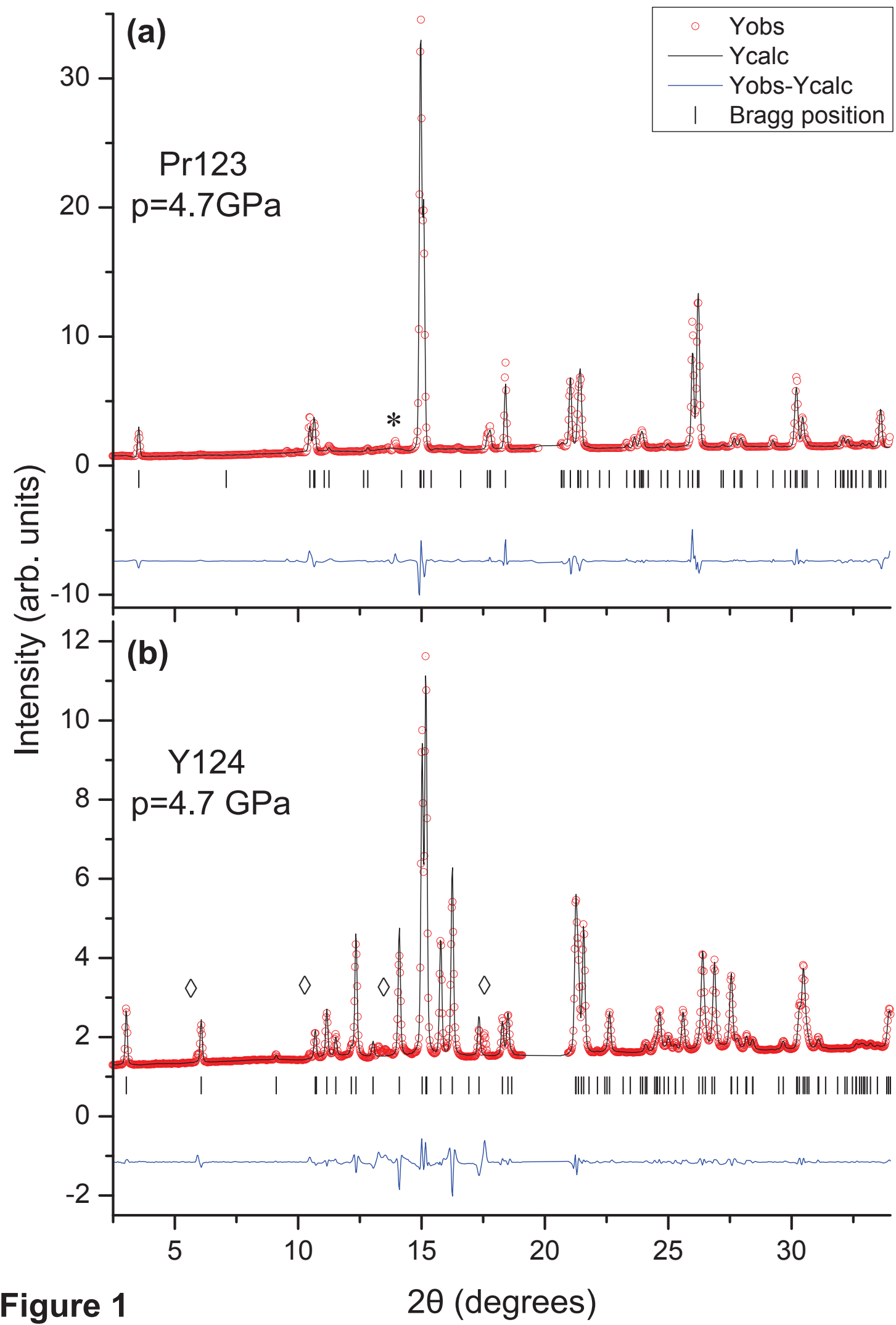
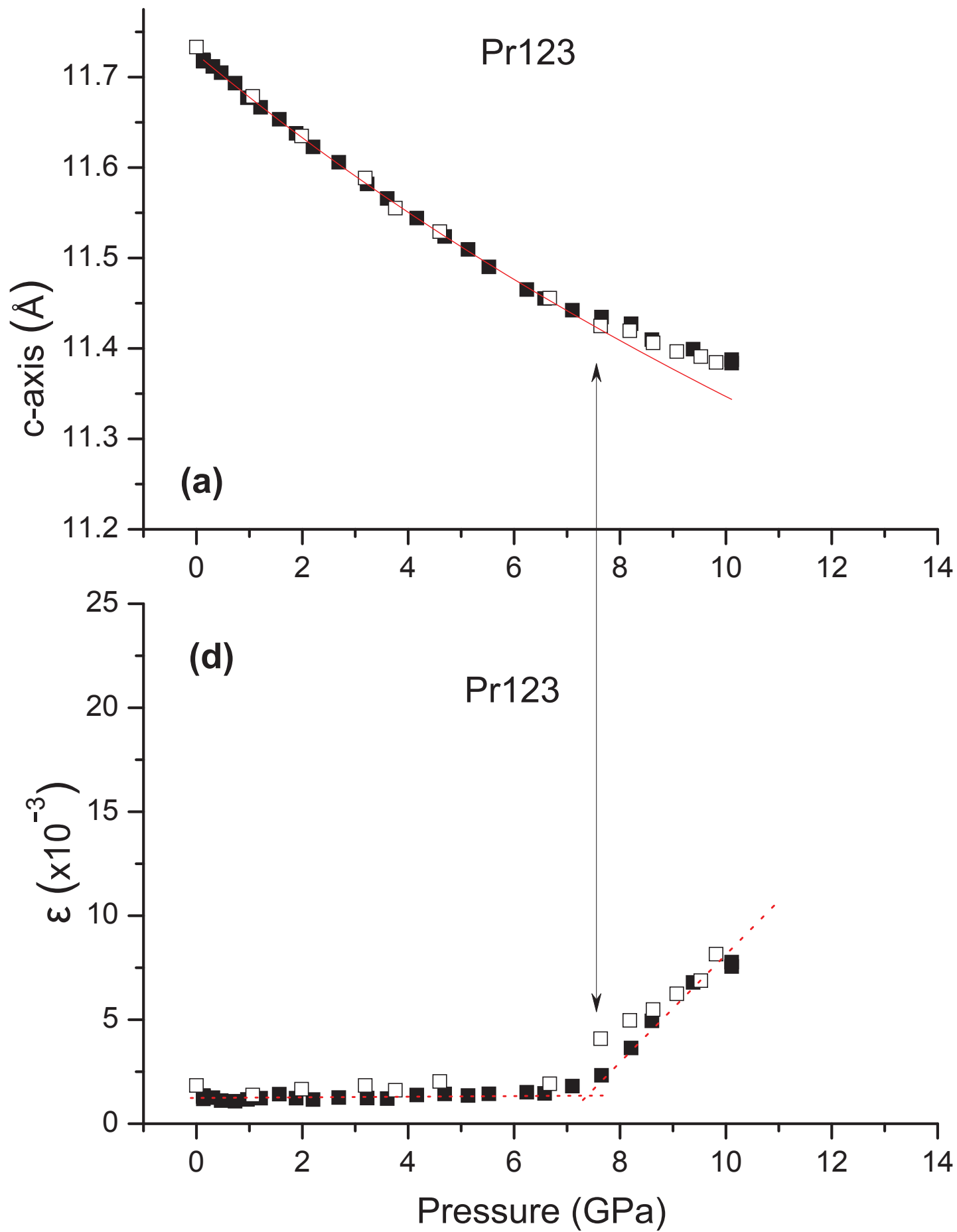
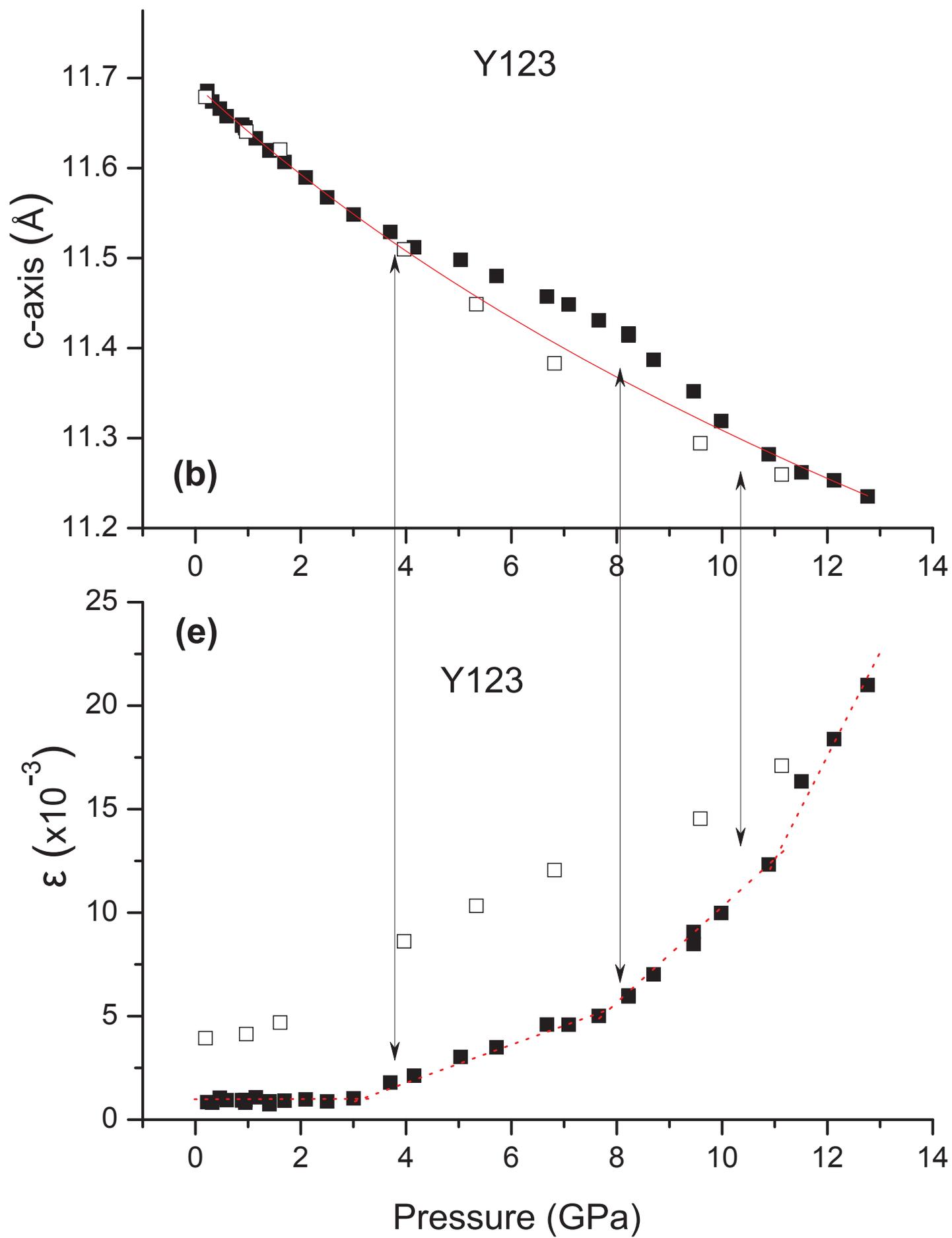
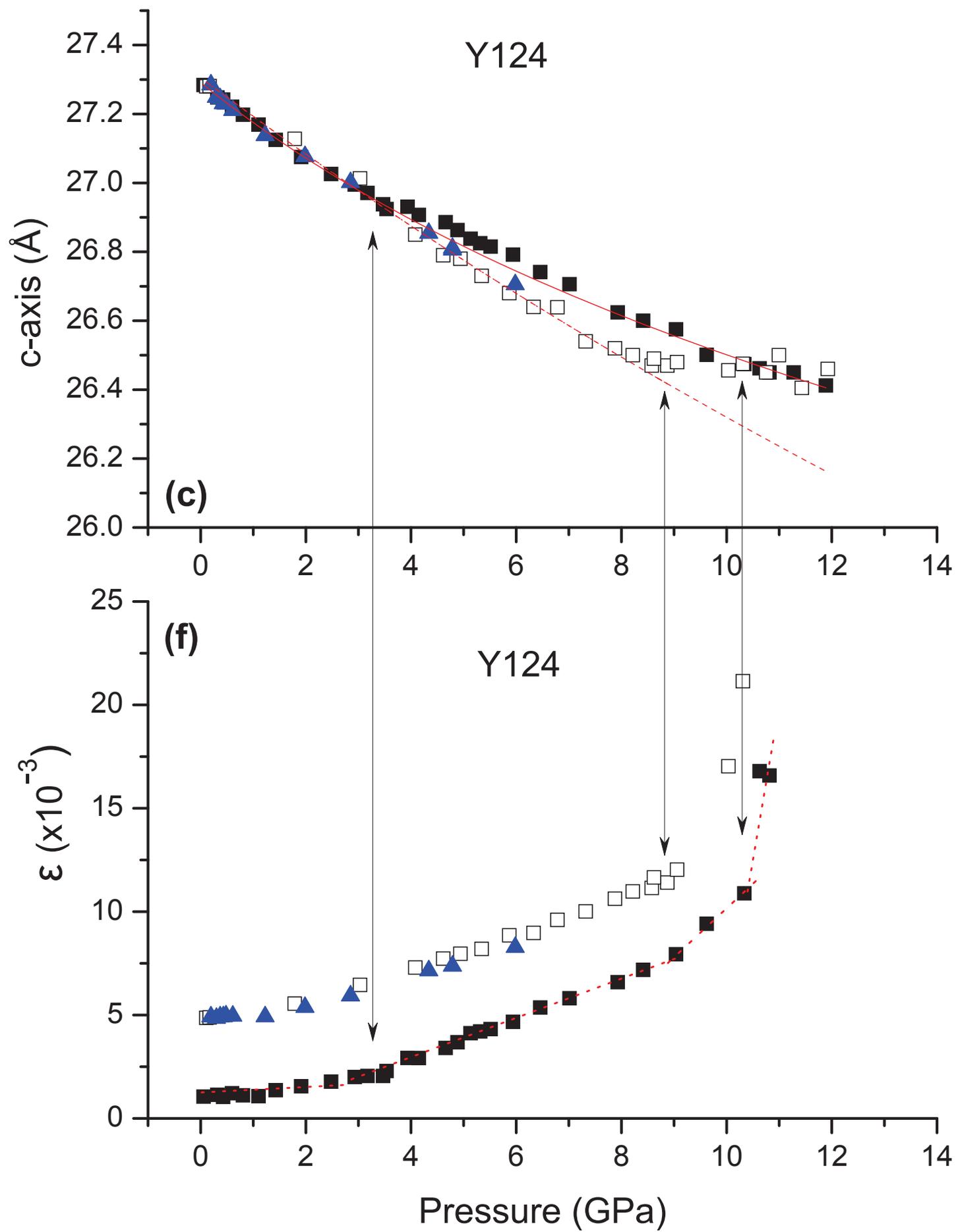
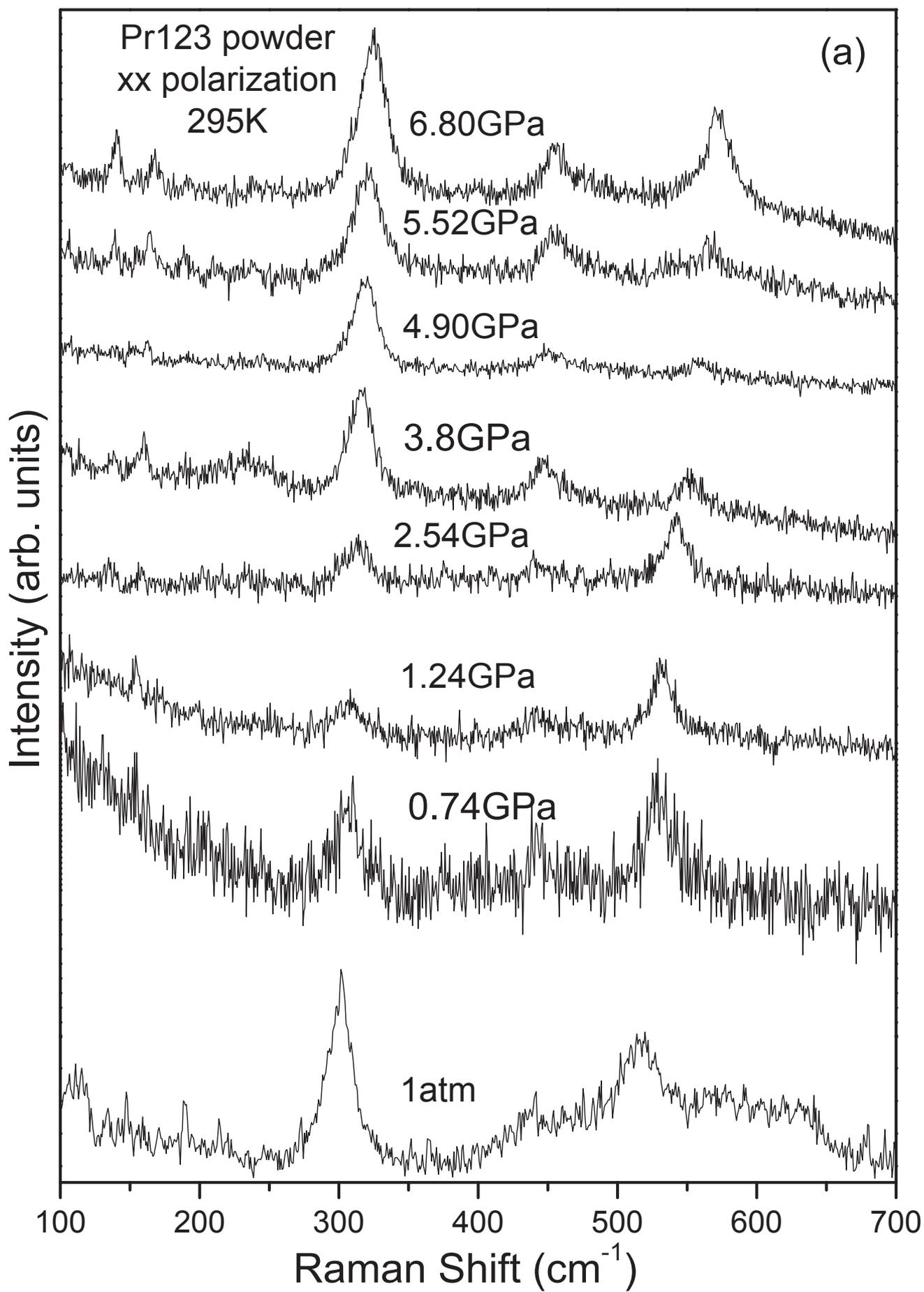
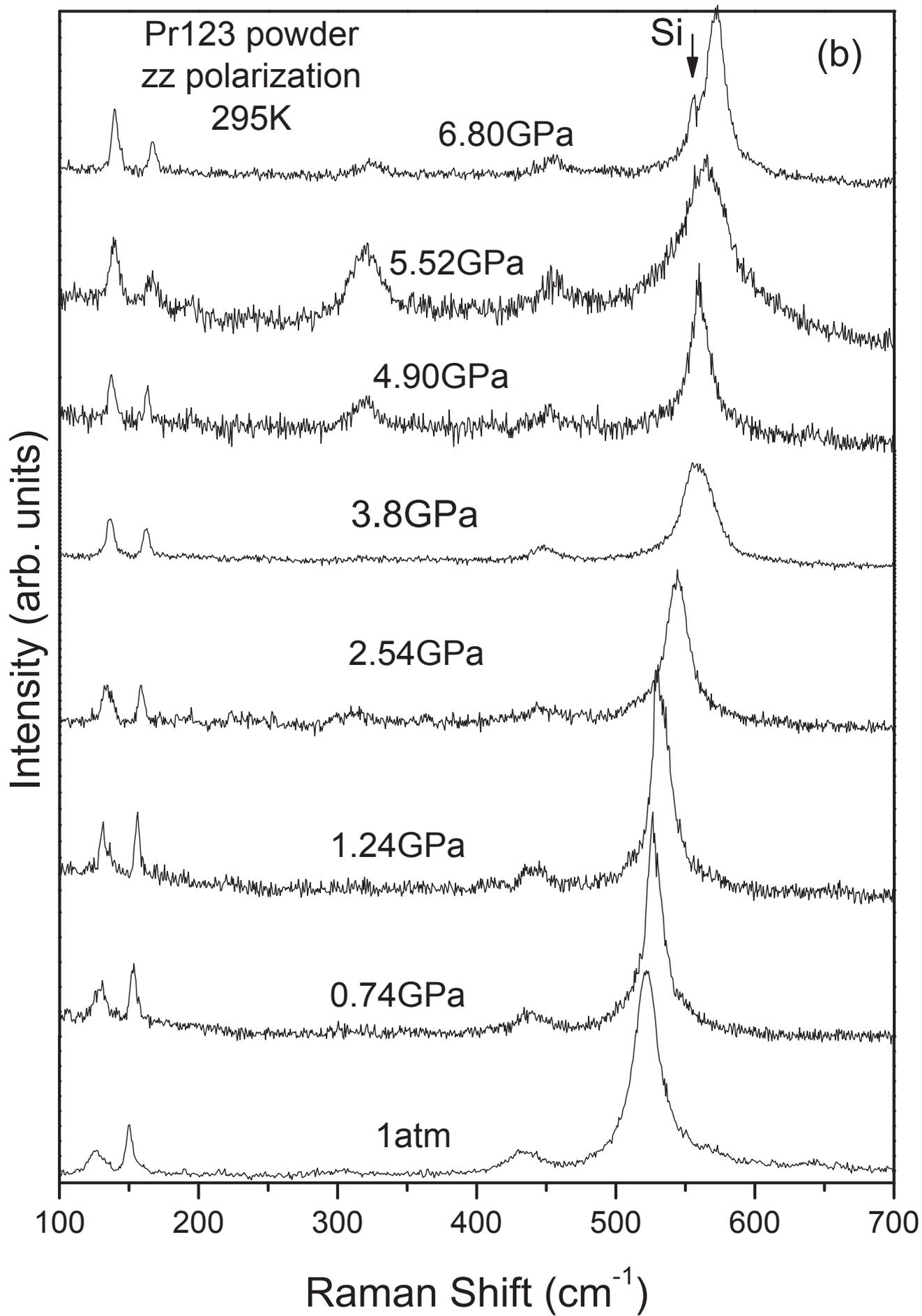
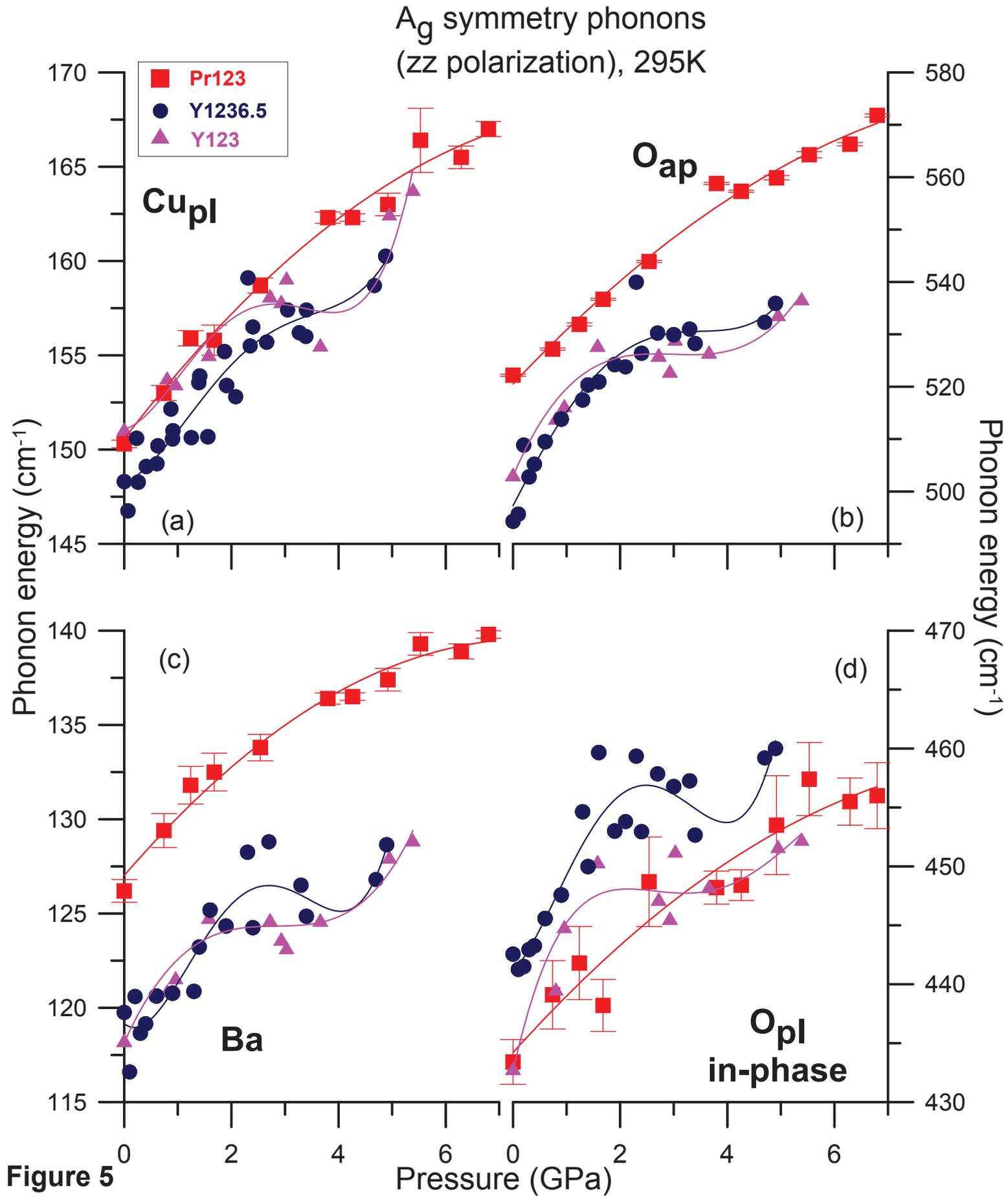
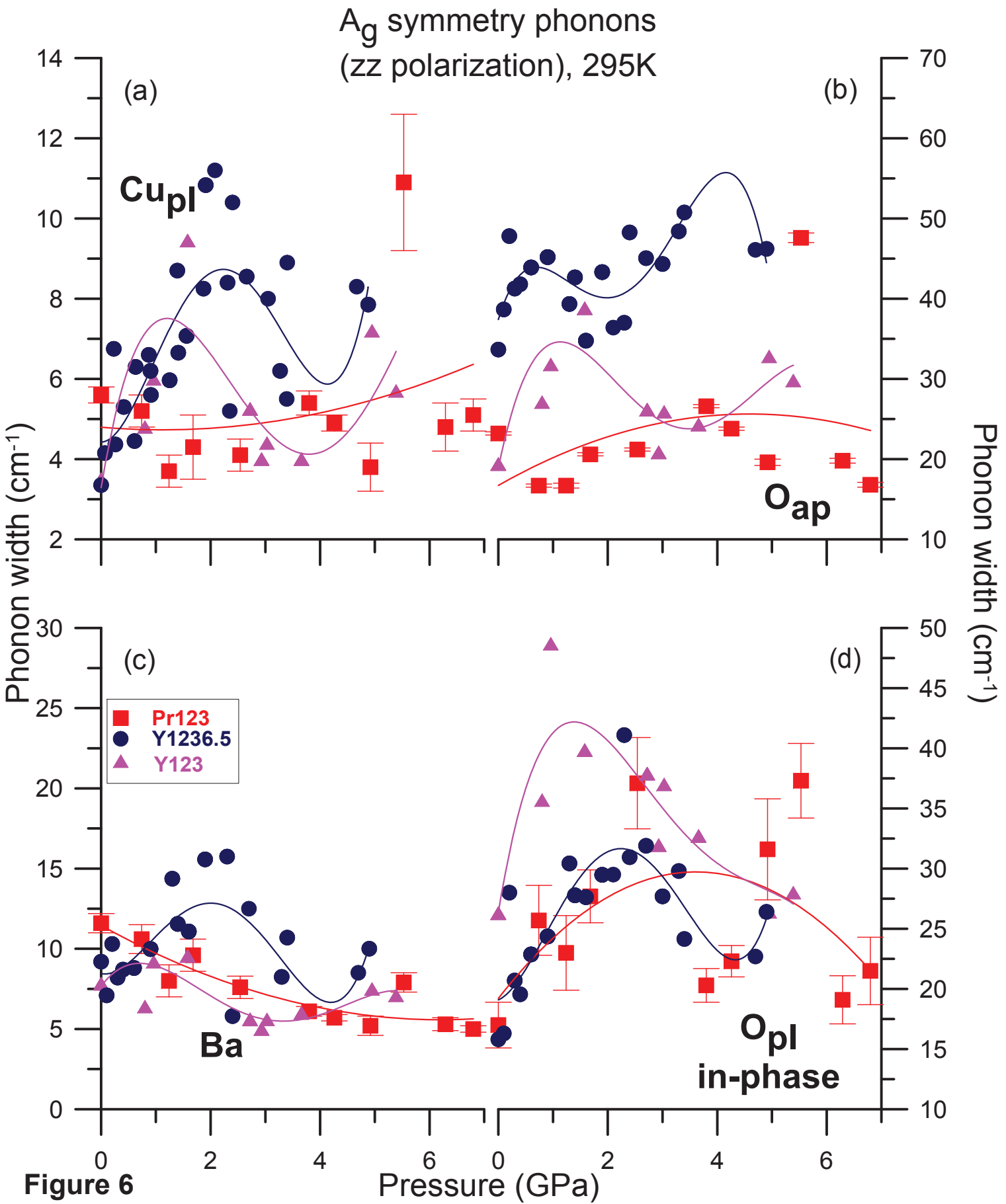
\end{document}